\documentclass[12pt]{article}

\usepackage{amssymb}
\usepackage{amsmath}
\usepackage{amscd}
\usepackage{latexsym}

\usepackage{graphicx}
\usepackage{array}

\usepackage{cite}

\newcommand{\be}{\begin{equation}}
\newcommand{\ee}{\end{equation}}
\newcommand{\Dlt}{\Delta}
\newcommand{\dlt}{\delta}

\newcommand{\br}{{\bf r}}
\newcommand{\bk}{{\bf k}}
\newcommand{\bn}{{\bf n}}
\newcommand{\bfe}{{\bf e}}

\newcommand{\bp}{{\bf p}}

\newcommand{\bt}{\beta}
\newcommand{\vp}{\varphi}

\newcommand{\al}{\alpha}
\newcommand{\ra}{\rightarrow}

\newcommand{\gm}{\gamma}
\newcommand{\om}{\omega}
\newcommand{\Om}{\Omega}

\newcommand{\bB}{{\bf B}}
\newcommand{\bS}{{\bf S}}

\newcommand{\rgl}{\rangle}
\newcommand{\lgl}{\langle}

\begin{document}

\begin{center}

{\Large{\bf Influence of quadratic Zeeman effect on spin waves in dipolar lattices} \\ [5mm]

V.I.~Yukalov$^{1,2,*}$ and E.P.~Yukalova$^{3}$ } \\ [3mm] 

{\it 
$^1$Bogolubov Laboratory of Theoretical Physics, \\
Joint Institute for Nuclear Research, Dubna 141980, Russia \\ [2mm]
                                           
$^2$Instituto de Fisica de S\~ao Carlos, Universidade de S\~ao Paulo, \\
CP 369,  S\~ao Carlos 13560-970, S\~ao Paulo, Brazil  \\ [2mm]

$^3$Laboratory of Information Technologies, \\
Joint Institute for Nuclear Research, Dubna 141980, Russia } \\ [1cm]

$^*$ Corresponding author
E-mail address:  yukalov@theor.jinr.ru (V.I. Yukalov)     

\end{center}

\vskip 1cm

\begin{abstract}

A lattice of particles with dipolar magnetic moments is considered under the presence
of quadratic Zeeman effect. Two types of this effect are taken into account, the effect
due to an external nonresonant magnetic field and the effect caused by an alternating 
quasiresonance electromagnetic field. The presence of the alternating-field quadratic 
Zeeman effect makes it possible to efficiently vary the sample characteristics. The 
main attention is payed to the study of spin waves whose properties depend on the 
quadratic Zeeman effect. By varying the quadratic Zeeman-effect parameter it is 
possible to either suppress or stabilize spin waves.        

\end{abstract}

\vskip 1cm

{\parindent =0pt


\vskip 2cm

{\it Declaration of interests}: none}

\newpage

\section{Introduction}

Spin waves is one of the important characteristics of magnetic materials, providing us
information on the latter. Also, spin waves have been found to be an essential ingredient 
in magnon spintronics \cite{Yu_32}. One usually studies spin waves in materials with 
magnetic exchange interactions, such as ferromagets, ferrimagnets, and antiferromagnets 
\cite{Akhiezer_33}. In these materials with self-organized magnetic order, spin waves can 
arise in the absence of external fields. While the existence of spin waves in materials with 
dipolar interactions requires a sufficiently strong external magnetic field.   

There exist many materials, whose constituents interact through magnetic dipolar forces. 
Such lattices can be formed by magnetic nanomolecules 
\cite{Kahn_1,Barbara_2,Caneschi_3,Yukalov_4,Yukalov_5,Yukalov_6},
magnetic nanoclusters \cite{Kodama_7,Hadjipanayis_8,Yukalov_9},
magnetic particles inserted into a nonmagnetic matrix \cite{Yukalov_10,Viali_11}. 
Different dipolar atoms and molecules can be arranged in self-assembled lattices 
or can form lattice structures with the help of superimposed external fields
\cite{Griesmaier_12,Baranov_13,Baranov_14,Gadway_15,Yukalov_16}. Many
biological systems contain molecules interacting through dipolar forces 
\cite{Cameretti_34,Waigh_35}. Numerous polymers are composed of nanomolecules
with dipolar interactions \cite{Barford_36}.

In some cases, the existence of spin waves in dipolar systems can be due, in addition 
to a sufficiently strong external magnetic field, to the arising quadratic Zeeman effect. 
It is important to distinguish two types of the latter.

One is the standard external-field quadratic Zeeman effect that appears, when atoms 
or molecules possess hyperfine structure \cite{Woodgate_17,Demtroder_18}. This 
effect, that has been described in extensive literature 
\cite{Jenkins_37,Schiff_38,Killingbeck_39,Coffey_40,Taylor_41}, induces in an atom 
with a dipole magnetic moment ${\bf m}_S$, the quadratic Zeeman energy shift
$$
\mp \; \frac{(\bB_0\cdot{\bf m}_S)^2}{\Dlt W(1+2I)^2} \;   ,
$$
proportional to the square of an external magnetic field ${\bf B}_0$, where $\Delta W$ 
is the hyperfine energy splitting, $I$ is nuclear spin, and the sign minus or plus corresponds 
to the parallel or, respectively, antiparallel alignment of the nuclear and total electronic spin 
projections in the atom. The effect can appear in any atom or molecule, whose nucleus 
possesses a nonzero spin. For example, in spinor atomic systems (not necessarily condensed) 
\cite{Yukalov_16,Stamper_26} and in magnetic admixtures inside nonmagnetic matrix  
\cite{Viali_11,Pajot_42,Veissier_43}. Quantum dots possess the properties similar to atoms 
and molecules \cite{Birman_44}, because of which quantum dots can also display quadratic 
Zeeman effect \cite{Prado_45}.

The other type of the effect is quasiresonance alternating-field quadratic Zeeman effect, 
due to the alternating-current Strak shift, induced either by applying off-resonance linearly 
polarized light exerting the quadratic shift along the polarization axis 
\cite{Cohen_19,Santos_20,Jensen_21,Paz_22} or by acting with a linearly polarized 
microwave driving field \cite{Gerbier_23,Leslie_24,Bookjans_25}. 

A linearly polarized microwave driving field generates the quasiresonance quadratic 
Zeeman effect by inducing hyperfine transitions in an atom  
\cite{Gerbier_23,Leslie_24,Bookjans_25}. While the alternating quasiresonance polarized 
light generates quadratic Zeeman effect by inducing transitions between the internal spin 
states and it exists even if the atom does not enjoy hyperfine structure  
\cite{Gerbier_23,Leslie_24,Bookjans_25}.

Both these alternating-current effects can be tailored with a high resolution and rapidly 
adjusted, producing a quadratic Zeeman shift described by the ratio $-\hbar\Om^2/4\Dlt$, 
where $\Omega$ is the driving Rabi frequency and $\Delta$ is the detuning from an internal
(spin or hyperfine) transition. This parameter does not depend on the external magnetic field
$B_0$ that is non-resonant with respect to internal atomic transitions, but depends only on
the intensity and detuning of the quasi-resonance driving field. By employing either positive 
or negative detuning, the sign of the above characteristic ratio can be varied. Below we accept
that the linear polarization of the alternating quasiresonance field is chosen along the 
$z$-axis.   

Thus, the alternating-current quasiresonant quadratic Zeeman effect can be realized in any 
atom, molecule, or quantum dot having nonzero spin. Therefore the parameters of the materials 
can be varied in a wide range. The theory and experimental realization of the quasiresonant 
quadratic Zeeman effect have been expounded in many publications, e.g., in
\cite{Cohen_19,Santos_20,Jensen_21,Paz_22,Gerbier_23,Leslie_24,Bookjans_25}. It is
important that this effect can be generated in atoms without hyperfine structure,
but merely having nonzero total electronic spins \cite{Cohen_19,Jensen_21,Paz_22}.

The quadratic Zeeman effect can be not very important for hard magnetic materials with
exchange interactions. But it can essentially influences the properties of dipolar materials,
whose constituents interact through dipolar forces that are much weaker than those due
to exchange interactions. This is why we study here dipolar lattices, whose characteristics 
can be regulated by quadratic Zeeman effect. The main attention is payed to the influence 
of this effect on spin waves. We show that by varying the parameters of the quadratic
Zeeman effect spin waves can be either suppressed or stabilized.

\section{Spin Hamiltonian}

The system Hamiltonian is a sum of two terms corresponding to the Zeeman, $\hat H_Z$,
and dipolar, $\hat H_D$, terms, 
\be
\label{1}
 \hat H = \hat H_Z + \hat H_D \;  .
\ee
These terms can be written in the following form  \cite{Yukalov_16,Stamper_26}.
The Zeeman part includes the linear Zeeman energy and quadratic Zeeman energy 
terms, 
\be
\label{2}
 \hat H_Z = \sum_j \left [ - \mu_S \bB_0 \cdot \bS_j + Q_Z ( \bB_0 \cdot \bS_j )^2
+ q_Z ( S_j^z )^2 \right ] \;  .
\ee
Here $\mu_S \equiv - g_S \mu_B$, with $g_S$ being the $g$-factor related to spin $S$
and $\mu_B$ is the Bohr magneton. $B_0$ is a static external magnetic field along the 
$z$-axis,
\be
\label{3}
 \bB_0 = B_0 \bfe_z \;  .
\ee

The parameter of the non-resonant magnetic-field induced Zeeman effect is
\be
\label{4}
 Q_Z = \mp \; \frac{\mu_S^2}{\Dlt W ( 1 + 2I)^2 } \;  ,
\ee
where $\Delta W$ is the hyperfine energy splitting and $I$ is nuclear spin. 

Alternating quasiresonance fields generate quadratic Zeeman effect either by acting 
on the atom by a linearly polarized microwave driving field inducing hyperfine transitions  
\cite{Gerbier_23,Leslie_24,Bookjans_25} or by applying off-resonance linearly polarized 
light populating internal spin states of the atom and inducing the quadratic Zeeman 
shift along the polarization axis \cite{Cohen_19,Santos_20,Jensen_21,Paz_22}. The 
last type of the effect exists even if the atom does not enjoy hyperfine structure. Both 
these methods can be rapidly adjusted, producing a quadratic Zeeman shift described 
by the parameter 
\be
\label{5}
 q_Z = -\; \frac{\hbar \Om^2}{4\Dlt} \;  ,
\ee
where $\Omega$ is the driving Rabi frequency and $\Delta$ is the detuning from an internal
(spin or hyperfine) transition. This parameter does not depend on the external  magnetic field
$B_0$ that is non-resonant with respect to internal atomic transitions, but depends only on
the intensity and detuning of the quasi-resonance driving field. By employing either positive 
or negative detuning, the sign of $q_Z$ can be varied. Here the linear polarization of the 
alternating quasiresonance field is chosen along the $z$-axis.   

Dipolar spin interactions are characterized  by the Hamiltonian
\be
\label{6}
 \hat H_D = \frac{1}{2} \; \sum_{i\neq j} \; \sum_{\al\bt} 
\overline D_{ij}^{\al\bt} S_i^\al S_j^\bt \;  ,
\ee
in which $\overline D_{ij}^{\al\bt}$ is a dipolar interaction potential. 

In some cases, one needs to take into account finite sizes of molecules and their mutual
correlations. Then, as suggested by Jonscher \cite{Jonscher_46,Jonscher_47,Jonscher_48},
it is possible to include the screening that can be characterized by an exponential function
\cite{Jonscher_46,Jonscher_47,Jonscher_48,Tarasov_49,Yukalov_27}. Therefore, for 
generality, we keep in mind the regularized dipolar interaction potential
\be
\label{7}
 \overline D_{ij}^{\al\bt} = \frac{\mu_S^2}{r_{ij}^3} \; \left ( \dlt_{\al\bt} -
3 n_{ij}^\al n_{ij}^\bt \right ) \exp(-\varkappa r_{ij} ) \;  ,
\ee
where
$$
 r_{ij} \equiv | \; \br_{ij}\; | \; , \qquad \bn_{ij} \equiv \frac{\br_{ij}}{r_{ij}} \; ,
\qquad \br_{ij} \equiv \br_i - \br_j \;  .
$$
Although this regularization is not principal for what follows. 

If necessary, one can estimate the interaction screening parameter $\varkappa = 1/r_s$, 
where $r_s$ is the screening radius, from the equality of the effective energy of spin 
interactions and of the effective kinetic energy, $\rho \mu_S^2 S^2 = \hbar^2/2m r_s^2$,   
where $\rho$ is average spin density. This gives
\be
\label{8}
 r_s = \frac{\hbar}{\sqrt{2m\rho \mu_S^2 S^2} } \;  .
\ee      
For example, if the spin density is $\rho \sim 10^{15}$ cm$^{-3}$, hence the mean interspin 
distance is $a \approx \rho^{-1/3} \sim 10^{-5}$ cm, and $\mu_S S \sim 10 \mu_B$, so that
the effective spin interaction energy is $\rho \mu_S^2 S^2 \sim 10^{-23}$ erg, then 
the screening radius $r_s \sim 10^{-5}$ cm is close to the mean distance $a$.  

But, as is stressed above, when one does not need to consider correlation effects, one 
can set $\varkappa$ to zero. So, for what follows the existence of screening is not important
and is kept only for generality. 

Thus the Zeeman Hamiltonian can be represented as
\be
\label{9}
 \hat H_Z = \sum_j \left [ - \mu_S B_0 S_j^z + Q ( S_j^z )^2 \right ] \;  ,
\ee
where the effective parameter of the quadratic Zeeman effect is
\be
\label{10}
 Q \equiv Q_Z B_0^2 + q_Z \;  .
\ee

Invoking the ladder operators $S^{\pm}_j = S^x_j \pm i S^y_j$, the dipolar Hamiltonian can be
written in the form
\be
\label{11}
 \hat H_D = \frac{1}{2} \sum_{i\neq j} \left [ a_{ij} ( S_i^z S_j^z \; - \;
\frac{1}{2} \; S_i^+ S_j^- ) + b_{ij} S_i^+ S_j^+ + b_{ij}^* S_i^- S_j^-
+ 2 c_{ij} S_i^+ S_j^z + 2 c_{ij}^* S_i^- S_j^z \right ] \;  ,
\ee
in which the notations are used:
$$
a_{ij} \equiv  \overline D_{ij}^{zz} = \frac{\mu_S^2}{r_{ij}^3} \; \left [ 1 - 3(n_{ij}^z)^2
\right ] \exp(-\varkappa r_{ij}) \; , 
$$
$$
b_{ij} \equiv \frac{1}{4} \; \left ( \overline D_{ij}^{xx} - \overline D_{ij}^{yy}
-2i \overline D_{ij}^{xy} \right ) = - \; \frac{3\mu_S^2}{4r_{ij}^3} \; 
\left ( n_{ij}^x - i n_{ij}^y \right )^2 \exp(-\varkappa r_{ij}) \; ,
$$
\be
\label{12}
 c_{ij} \equiv \frac{1}{2} \; \left ( \overline D_{ij}^{xz} - i\overline D_{ij}^{yz} \right )
= - \; \frac{3\mu_S^2}{2r_{ij}^3} \; \left ( n_{ij}^x - i n_{ij}^y \right ) n_{ij}^z
 \exp(-\varkappa r_{ij}) \;  .
\ee
At short distance, dipolar interactions enjoy the natural conditions excluding self-interactions,  
$$
 \overline D_{jj}^{\al\bt} \equiv 0 \; , \qquad 
a_{jj} \equiv b_{jj} \equiv c_{jj} \equiv 0 \; .
$$

For a large lattice, where the boundary effects can be neglected, one has
\be
\label{13}
\sum_j \overline D_{ij}^{\al\bt} = 0\;   ,
\ee
hence the interaction terms (\ref{12}) satisfy the equalities
$$
 \sum_j a_{ij} = \sum_j b_{ij} = \sum_j c_{ij} = 0 \; .
$$

The quantities
$$
\xi_i \equiv \frac{1}{\hbar} 
\sum_j \left ( a_{ij} S_j^z + c_{ij} S_j^+ + c_{ij}^* S_j^- \right ) \; ,
$$
\be
\label{14}
\vp_i \equiv \frac{1}{\hbar} 
\sum_j \left ( \frac{a_{ij}}{2} \; S_j^- - 2 b_{ij} S_j^+ -2 c_{ij} S_j^z \right ) 
\ee
play the role of the local fields acting on spins. For an ideal lattice, these quantities 
are zero centered, so that 
\be
\label{15}
 \lgl \xi_j \rgl = \lgl \vp_j \rgl = 0 \;   ,  
\ee
which follows from equation (\ref{13}).

\section{Spin waves}

The equations of motion for the spin operators read as
$$
\frac{d S_j^-}{dt} = - i ( \om_0 + \xi_j ) S_j^- - i\vp_j S_j^z - i\; \frac{Q}{\hbar} \;
\left ( S_j^- S_j^z + S_j^z S_j^- \right ) \; ,
$$
\be
\label{16}
\frac{d S_j^z}{dt} = - \; \frac{i}{2} \; \left ( \vp_j^+ S_j^- - S_j^+ \vp_j \right ) \; ,
\ee
where the Zeeman frequency is
\be
\label{17}
 \om_0 \equiv -\; \frac{\mu_S B_0}{\hbar} > 0 \; .
\ee

Spin waves are defined as small spin fluctuations around the average spin values
$\langle S_j^\alpha \rangle$. This average can correspond to a stationary or 
quasistationary state. Quasistationary are the states with lifetime longer then the 
oscillation time $2 \pi / \omega_0$. Following the standard technique \cite{Tyablikov_28}, 
we represent the spin operators in the form
\be
\label{18}
 S_j^\al = \lgl S_j^\al \rgl + \dlt S_j^\al \;  ,
\ee
in which $\delta S_j^\alpha$ is a small deviation from the average $\lgl S_j^\alpha \rgl$. 
In the stationary state, the spins are assumed to be directed along the field ${\bf B}_0$, 
that is, along the $z$-axis, so that
\be
\label{19}
\lgl S_j^\pm \rgl = 0\; , \qquad \lgl S_j^z \rgl \neq 0 \; .
\ee
Then the local fields (\ref{14}) become
$$
\xi_i \equiv \frac{1}{\hbar} 
\sum_j \left ( a_{ij} \dlt S_j^z + c_{ij} \dlt S_j^+ + 
c_{ij}^* \dlt S_j^- \right ) \; , 
$$
\be
\label{20}
\vp_i \equiv \frac{1}{\hbar} 
\sum_j \left ( \frac{a_{ij}}{2} \;\dlt S_j^- - 2 b_{ij} \dlt S_j^+ 
-2 c_{ij} \dlt S_j^z \right ) \;  .
\ee
  
Representation (\ref{18}) is substituted into equations (\ref{16}) that are linearized
with respect to the deviations $\delta S_j^\alpha$, taking into account that 
$S_j^- = \delta S_j^-$, according to equations (\ref{18}) and (\ref{19}). For the 
single-site expression in the right-hand site of the first of equations (\ref{16}),
we use the form
\be
\label{21}
S_j^- S_j^z + S_j^z S_j^- = \left ( 2 \; - \; \frac{1}{S} \right )
\lgl S_j^z \rgl S_j^-
\ee
that is exact for spin one-half and is asymptotically exact for spin $S \ra \infty$,
as is explained in Refs. \cite{Yukalov_4,Yukalov_5,Yukalov_29}. Thus we come 
to the equations
\be
\label{22}
 \frac{d}{dt} \; S_j^- = - \om_s S_j^- - i\vp_j \lgl S_j^z \rgl \; , \qquad
\frac{d}{dt}\dlt S_j^z = 0 \;  ,
\ee
in which
\be
\label{23}
\om_s \equiv \om_0  + \left ( 2 \; - \; \frac{1}{S} \right ) \; \frac{Q}{\hbar} \; 
\lgl S_j^z \rgl
\ee
is the effective frequency of spin rotation. With the initial condition $\dlt S_j^z(0)=0$,
one has $\delta S_j^z(t) = 0$. Then the local fields (\ref{20}) take the form
\be
\label{24}
\xi_i \equiv \frac{1}{\hbar} \sum_j \left ( c_{ij}  S_j^+ + c_{ij}^*  S_j^- \right ) \; , 
\qquad
\vp_i \equiv \frac{1}{\hbar} 
\sum_j \left ( \frac{a_{ij}}{2} \; S_j^- - 2 b_{ij} S_j^+ \right ) \;  .
\ee

Let us define the Fourier transforms for the spin operators,
\be
\label{25}
S_j^\pm = \sum_k S_k^\pm \exp (\mp i\bk \cdot \br_j ) \; , \qquad
S_k^\pm = \frac{1}{N} \sum_j S_j^\pm \exp (\pm i\bk \cdot \br_j ) 
\ee
and for the interaction terms
\be
\label{26}
 a_{ij} = \frac{1}{N} \sum_k a_k \exp (i\bk \cdot \br_{ij} ) \; , \qquad
 a_k =  \sum_j a_j \exp (- i\bk \cdot \br_{ij} ) \;.
\ee
The Fourier transform for $b_{ij}$ is defined similarly to Eq. (\ref{26}). 

In this way, we come to the equation
\be
\label{27}
 \frac{d}{dt} \; S_k^- = - i A_k S_k^- + i B_k S_k^+ \;  ,
\ee
where
\be
\label{28}
  A_k \equiv \om_s + \frac{a_k}{2\hbar} \; \lgl S_j^z \rgl \; , \qquad
B_k \equiv \frac{2b_k}{\hbar} \; \lgl S_j^z \rgl \;  .
\ee
We may notice that $A_k$ is real, since $a_{ij} = a_{ji} = a_{ij}^*$. 

Looking for the solution in the form
$$
 S_k^- = u_k e^{-i\om_k t} + v_k^* e^{i\om_k t} \;  ,
$$
we get the eigenvalue equations
$$
 A_k u_k - B_k v_k = \om_k u_k \; , \qquad  B_k^* u_k - A_k v_k = \om_k v_k \; .
$$
From here we find the spectrum of spin waves
\be
\label{29}
  \om_k = \sqrt{A_k^2 - |\; B_k \; |^2 } \;  .
\ee
In the long-wave limit, the spectrum is quadratic,
\be
\label{30}
 \om_k \simeq | \; \om_s \; | \left [ 1 \; - \; \frac{\lgl S_j^z\rgl }{4\hbar\om_s} \;
\sum_{\lgl ij\rgl } a_{ij} ( \bk\cdot\br_{ij} )^2 \right ] \;  ,
\ee
where $k \ra 0$ and the summation is over the nearest neighbors.

\section{Cubic lattice}

For concreteness, let us consider a cubic lattice with the side $a$. Then for each 
lattice site there are six nearest neighbors, so that the unit vector ${\bf n}_{ij}$ for six 
values of $j$, corresponding to the nearest neighbors to a site $i$, has the following
components
$$
n_{ij}^x = \{ 1,\; -1,\; 0, \; 0, \; 0, \; 0 \} \; , \qquad
n_{ij}^y = \{ 0,\; 0,\; 1, \; -1, \; 0, \; 0 \} \; , 
$$
\be
\label{31}
n_{ij}^z = \{ 0,\; 0,\; 0, \; 0, \; 1, \; -1 \}  \; .
\ee
Then the Fourier transforms for the interaction terms, defined in Eq. (\ref{26}), become
$$
a_k = 2\rho \mu_S^2 [ \; \cos(k_x a) + \cos(k_y a) - 2 \cos(k_z a) \; ] \; , 
$$
\be
\label{32}
b_k = -\; \frac{3}{2} \; \rho \mu_S^2 [ \; \cos(k_x a) - \cos(k_y a)  \; ] \;  ,
\ee
and $c_k = 0$. 

It is convenient to pass to the dimensionless expression of the spin-wave spectrum
\be
\label{33}
\om(\bp ) \equiv \frac{\om_k}{\om_0 } 
\ee
that is a function of the dimensionless momentum
\be
\label{34}
\bp \equiv \bk a \qquad ( p_\al \equiv k_\al a) \;   .
\ee
Also, let us introduce the dimensionless parameter of the quadratic Zeeman effect
\be
\label{35}
\zeta \equiv \frac{Q}{\hbar\om_0} 
\ee
and the dimensionless strength of dipolar interactions
\be
\label{36}
\gm_D \equiv \frac{\rho\mu_S^2}{\hbar\om_0}   .
\ee
For the spin-rotation frequency (\ref{23}), we have
\be
\label{37}
\frac{\om_s}{\om_0} = 1 + \left ( 2 \; - \; \frac{1}{S} \right ) \zeta S_0 \;  ,
\ee
where
\be
\label{38} 
 S_0 \equiv \lgl S_j^z \rgl \;  .
\ee
And instead of $A_k$ and $B_k$, we introduce the dimensionless quantities
$$
\al_p \equiv \frac{A_k}{\om_0} = 1 + \left ( 2 \; - \; \frac{1}{S} \right ) \zeta S_0
+ \gm_D S_0 ( \cos p_x + \cos p_y - 2\cos p_z ) \; ,
$$
\be
\label{39}
 \bt_p \equiv \frac{B_k}{\om_0} = -3\gm_D S_0 ( \cos p_x - \cos p_y ) \; .
\ee

The eigenvalue equations for the spin-wave spectrum take the form
\be
\label{40}
 \al_p u_p - \bt_p v_p = \om(\bp) u_p \; , \qquad
 \bt_p u_p - \al_p v_p = \om(\bp) v_p \; ,
\ee
which gives the spectrum
\be
\label{41}
 \om(\bp) = \sqrt{\al_p^2 - \bt_p^2 } \;  .
\ee

When the momentum ${\bf p}$ is along the external magnetic field, such that 
${\bf p} = p_z {\bf e}_z$, then $\beta_p = 0$ and the eigenvalue equations (\ref{40})
reduce to 
$$
\al_p u_p  = \om(\bp) u_p \; , \qquad    \al_p v_p  = - \om(\bp) v_p \; .
$$
These equations do not possess nontrivial solutions for $u_p$ and $v_p$, which
implies that spin waves do not propagate along the direction of the external magnetic 
field ${\bf B}_0$. 

For the transverse propagation with respect to the external magnetic field, we can set
\be
\label{42}
 p_x = p \; , \qquad p_y = p_z = 0 \; .
\ee
Then functions (\ref{39}) read as
\be
\label{43}
\al_p =  1 + \left ( 2 \; - \; \frac{1}{S} \right ) \zeta S_0 - \gm_D S_0 ( 1 - \cos p ) \; , 
\qquad
\bt_p = 3 \gm_D S_0 ( 1 - \cos p ) \; .
\ee
The long-wave limit of the spin-wave spectrum (\ref{41}) is
\be
\label{44}
 \om(\bp) \simeq |\; C \; | \; \left ( 1 \; - \; \frac{\gm_D S_0}{2C} \; p^2 \right ) 
\qquad ( p \ra 0 ) \; ,
\ee
where
\be
\label{45}
 C \equiv 1 + \left ( 2 \; - \; \frac{1}{S} \right ) \zeta S_0 \; .
\ee

\section{Stability conditions}

A well defined spectrum of stable spin waves presupposes that it is non-negative:
\be
\label{46}
\om(\bp) \geq 0 \;   .
\ee
Otherwise, when it is complex, spin waves are not stable, but decay. Thus, if the 
expression under the square root in Eq. (\ref{41}) is negative, then the spectrum becomes 
imaginary, such that $\omega_k = i |\omega_k|$. Then spin waves are described by the 
operator
$$
 S_k^- = u_k e^{|\om_k|t} + v_k^*  e^{- |\om_k|t} 
$$    
showing that the spin-wave stability is lost after the time $2 \pi/ \omega_k$.
 
To be defined as a real quantity, spectrum (\ref{41}) requires that the expression under 
the square root be non-negative, which implies the stability condition
\be
\label{47}
 \left ( C \; - \; \frac{4}{3} \; \bt_p \right ) 
\left ( C \; + \; \frac{2}{3} \; \bt_p \right ) \geq 0 \; .
\ee

For a more detailed investigation of stability, we need to specify the average spin $S_0$. 
In the state of absolute equilibrium, the latter is defined from the minimization of the system 
free energy. More generally, $S_0$ can be prepared by polarizing the system at the initial 
moment of time and then considering the system behavior. Such a setup with a prepared 
polarization is very important for studying spin dynamics from an initially prepared state 
\cite{Yukalov_4,Yukalov_5,Yukalov_29}. The spin motion from a prepared initial state is 
triggered by spin waves, because of which the existence of the latter plays a crucial role 
for spin dynamics. There are two opposite cases of initial polarization. One corresponds 
to an initially polarized state with a positive polarization $S_0 > 0$, while the second, 
to an equilibrium state with a negative polarization $S_0 < O$. We shall consider both 
these cases.
    
If the average spin polarization is positive, hence $\beta_p$ is non-negative, then
the stability condition (\ref{47}) is valid when either
\be
\label{48}
C \geq \frac{4}{3} \; \bt_p \qquad ( \bt_p \geq 0 ) \; , 
\ee
or when 
\be
\label{49}
C \leq -\; \frac{2}{3} \; \bt_p \qquad ( \bt_p \geq 0 ) \;  .
\ee
Since these inequalities have to be valid for all $p \in [-\pi, \pi]$, they reduce to the 
conditions requiring that either
\be
\label{50}
C \geq 8 \gm_D S_0 \qquad ( S_0 > 0 ) \; ,
\ee
or
\be
\label{51}
 C \leq -4 \gm_D S_0 \qquad ( S_0 > 0 ) \;  .
\ee
In dimensional units, this means that either
\be
\label{52}
\hbar \om_0 + \left ( 2 \; - \; \frac{1}{S} \right ) Q S_0 \geq 8 \rho \mu_S^2 S_0 \; ,
\ee
or
\be
\label{53}
 \hbar \om_0 + \left ( 2 \; - \; \frac{1}{S} \right ) Q S_0 \leq -4 \rho \mu_S^2 S_0 \; ,
\ee
where $S_0 > 0$. This should be compared with the condition of stability for the 
case when the quadratic Zeeman effect is absent,
\be
\label{54}
\hbar \om_0 \geq 8 \rho \mu_S^2 S_0 \qquad ( Q = 0 , \; S_0 > 0 ) \; .
\ee
The latter condition means that a sufficiently strong external magnetic field, that is 
much larger than the effective strength of dipolar interactions, stabilizes spin waves. 
These cannot exist in dipolar systems without such a strong external field. 

The existence of the quadratic Zeeman effect extends the region of the magnetic-field 
strength, where spin waves are stable. The external magnetic field can be very small,
although spin waves perfectly exist, provided that the quadratic Zeeman parameter $Q$
is sufficiently large and positive in case (\ref{52}) or sufficiently large by its 
magnitude and negative in case (\ref{53}).  In that sense, the quadratic Zeeman effect 
stabilizes spin waves. For example, it may happen that condition (\ref{54}) does not 
hold, hence spin waves do not arise in the absence of the quadratic Zeeman effect. But 
switching on the quadratic Zeeman effect condition (\ref{52}) may become valid. Then 
spin waves can exist as stable collective excitations. On the contrary, when condition 
(\ref{54}) holds true, spin waves exist without the quadratic Zeeman effect. Then 
switching on this effect corresponding to a negative $Q$ can lead to the situation 
when neither condition (\ref{52}) nor (\ref{53}) are satisfied. This means that spin 
waves become suppressed. 

The other situation occurs, if the stationary spin polarization is negative, $S_0 < 0$,
hence $\beta_p$ is nonpositive. In such a case, spin waves are stable if either
\be
\label{55}
C \geq \frac{2}{3} \; | \;\bt_p \; | \qquad ( \bt_p \leq 0 ) 
\ee
or
\be
\label{56}
 C \leq -\;\frac{4}{3} \; | \;\bt_p \; | \qquad ( \bt_p \leq 0 )   .
\ee
To be valid for all $p \in [-\pi, \pi]$, these conditions result in the validity of the 
inequality
\be
\label{57}
C \geq 4\gm_D |\; S_0 \; | \qquad ( S_0 < 0 ) 
\ee
or, respectively,
\be
\label{58}
C \leq -8 \gm_D |\; S_0 \; | \qquad ( S_0 < 0 )   .
\ee
In dimensional units, we have either condition 
\be
\label{59}
\hbar \om_0 - \left ( 2 \; - \; \frac{1}{S} \right ) Q |\; S_0\; |  \geq 
4 \rho \mu_S^2 |\; S_0\; | 
\ee
or
\be
\label{60}
 \hbar \om_0 - \left ( 2 \; - \; \frac{1}{S} \right ) Q |\; S_0\; |  \leq 
-8 \rho \mu_S^2 |\; S_0\; | \; .
\ee
And if the quadratic Zeeman effect is absent, then spin waves are stable if
\be
\label{61}
 \hbar \om_0 \geq  4 \rho \mu_S^2 |\; S_0\; | \qquad ( Q = 0 , \; S_0 < 0 ) \; .
\ee
Again we see that, depending on the values of the Zeeman frequency $\omega_0$
and the quadratic Zeeman effect parameter $Q$, this effect can either stabilize of 
suppress spin waves. 
 
The conditions of stability for spin waves with respect to the value of the quadratic
Zeeman effect parameter $Q$ are summarized as follows: For a positive polarization,
spin waves are stable provided that either
\be
\label{62}
Q \geq -\; \frac{\hbar\om_0-8\rho\mu_S^2 S_0}{2S-1} \; \left ( \frac{S}{S_0} \right )
\qquad ( S_0 > 0 ) 
\ee
or
\be
\label{63}
Q \leq -\; \frac{\hbar\om_0+4\rho\mu_S^2 S_0}{2S-1} \; \left ( \frac{S}{S_0} \right )
\qquad ( S_0 > 0 ) \;  .
\ee
And in the case of a negative polarization, spin waves are stable when either
\be
\label{64}
Q \leq \frac{\hbar\om_0-4\rho\mu_S^2|S_0|}{2S-1} \; \left ( \frac{S}{ |S_0| } \right )
\qquad ( S_0 < 0 ) 
\ee
or
\be
\label{65}
Q \geq \frac{\hbar\om_0+8\rho\mu_S^2|S_0|}{2S-1} \; \left ( \frac{S}{ |S_0| } \right )
\qquad ( S_0 < 0 ) \; .
\ee

Recall that the quadratic Zeeman effect parameter $Q$, as defined in Eq. (\ref{10}),
consists of a nonresonant field term and of an alternating quasiresonance field term.
The latter can be varied in a rather wide range, because of which the parameter $Q$
is also changeable. In that way, by varying this parameter $Q$, one can either stabilize
spin waves or suppress them. 

To be more specific, let us consider the case of spin $S = 1$, when notation (\ref{45})
reduces to
\be
\label{66}
 C = 1 + \zeta S_0 \qquad ( S = 1 ) \;  .
\ee
Setting the spin polarization to be positive $S_0 = 1$, we find that spin waves are stable
when either
\be
\label{67}
\zeta \geq -1 + 8 \gm_D \qquad ( S_0 = 1 ) 
\ee
or
\be
\label{68}
 \zeta \leq -1 -4 \gm_D \qquad ( S_0 = 1 ) \;  .
\ee
While in the case of the negative polarization $S_0 = -1$, the conditions of spin wave 
stability become either
\be
\label{69}
 \zeta \leq 1 -4 \gm_D \qquad ( S_0 = - 1 ) 
\ee
or
\be
\label{70}
 \zeta \geq 1 +8 \gm_D \qquad ( S_0 = - 1 ) \;  .
\ee
Here $\zeta$ is the dimensionless quadratic Zeeman effect parameter (\ref{35}) and
$\gamma_D$ is the dimensionless strength of dipolar interactions (\ref{36}).

\section{Spin-wave spectrum} 

To illustrate how the spin-wave spectrum is influenced by the quadratic Zeeman effect,
we present below numerical calculations for spectrum (\ref{41}) corresponding to the 
transverse propagation of spin waves, with the dimensionless momentum (\ref{42}).   

Figure 1 demonstrates the spin-wave spectrum $\omega(p)$, under the positive 
polarization of the average spin $S_0 = 1$ and the dipolar parameter $\gamma_D = 0.1$,
for different parameters (\ref{35}) of the quadratic Zeeman effect. The spectrum is
stable if either $\zeta \geq -0.2$ (Fig.1a) or $\zeta \leq -1.4$ (Fig. 1b).  In Fig. 1a, 
if the quadratic Zeeman effect is switched off, the spectrum has a gap. Positive values 
of $\zeta$ shift the spectrum up, while negative values of $\zeta$ move it down. 
At a value of $\zeta = -0.2$, the gap disappears. If the value of $\zeta$ is diminished 
further below $-0.2$, the spectrum becomes imaginary, hence spin waves become 
unstable. But for $\zeta \leq -1.4$ the spectrum again stabilizes, which is shown in
Fig. 1b. Here the spectrum with $\zeta = -1.4$ is gapless, while decreasing $\zeta$
below $-1.4$ shifts the spectrum up and makes it gapful. 
 
Figure 2 presents the spectrum of spin waves under negative average spin polarization
$S_0 = -1$, for the strength of dipolar interactions $\gamma_D = 0.1$, and different
quadratic Zeeman-effect parameters $\zeta$. The spectrum is stable for $\zeta \geq 1.8$
(Fig. 2a) or $\zeta \leq 0.6$ (Fig. 2b). Varying the parameter $\zeta$ one can make the 
spectrum gapful or gapless. 

The feasibility of influencing the properties of spin waves by the quadratic Zeeman effect
can be used for regulating spin dynamics. From equation (\ref{27}), one sees that the
rotation speed of the average spin is influenced by the quadratic Zeeman-effect parameter
$Q$. When the spin system is prepared in an initial nonequilibrium (or quasiequilibrium) state,
then the velocity of spin motion essentially depends on the strength and the oscillation 
frequency of spin waves that serve as a trigger for starting the spin dynamics 
\cite{Yukalov_30,Yukalov_31}. The possibility of regulating spin dynamics can be employed 
in spintronics and in quantum information processing.

\section{Conclusion}

We have considered a dipolar lattice subject to the action of the usual linear Zeeman effect
and also of the quadratic Zeeman effect. The latter can be of two types, the constant-field
quadratic Zeeman effect and the alternating-current quadratic Zeeman effect. Both these 
cases are taken into account. The existence of the quadratic Zeeman effect can strongly 
influence the properties of spin waves. The feasibility of regulating the strength of this
influence makes it possible to vary the spectrum of spin waves and their stability. Since 
spin waves serve as a triggering mechanism initiating spin rotation in spin systems prepared
in a nonequilibrium state, the regulation of spin-wave properties can be used as a tool for 
governing spin dynamics in spintronics and in quantum information processing. This problem 
of spin dynamics requires a separate investigation and will be done in a separate paper. 

As is explained in the Introduction, there exists plenty of atoms or molecules interacting 
through dipolar forces and possessing quadratic Zeeman effect. It is therefore possible
to vary the system parameters in a very wide range. In order to illustrate by a particular
example that the quadratic Zeeman effect can really be sufficiently large, such that it
would be feasible to use it for regulating the properties of spin waves, caused by dipolar
interactions, let us consider the case of $^{52}$Cr. This case is interesting, since the 
nuclear spin of this atom is zero, so that $^{52}$Cr does not have hyperfine structure, because 
of which the stationary-field parameter defined in Eq. (\ref{4}), is $Q_Z = 0$. And the 
alternating-field parameter of the quadratic Zeeman effect, defined in Eq. (\ref{5}) can be made
\cite{Santos_20} as large as $q_Z \sim 10^5 \hbar/ {\rm s}$. Taking for typical dipolar lattices
\cite{Yukalov_16,Gadway_50} the density of atoms $\rho \sim (10^{12} - 10^{15})$ cm$^{-3}$,
and the dipolar magnetic moment $\mu_S \sim (1 - 10) \mu_B$, where $\mu_B$ is the Bohr 
magneton, we get $\rho \mu_S^2 \sim (1 - 10^4) \hbar / {\rm s}$. Then the ratio of the 
quadratic Zeeman parameter (\ref{35}) to the strength of dipolar interactions (\ref{36}) is
$$
 \frac{\zeta}{\gm_D} = \frac{q_Z}{\rho\mu_S^2} \sim 10 - 10^5 \;  .
$$
Hence the quadratic Zeeman effect can essentially influence the properties of spin waves,
either suppressing or stabilizing them.

\newpage

\begin{center}
{\bf {\Large Figure Captions} }

\end{center}

\vskip 1cm
{\bf Figure 1}. Dimensionless spectrum of spin waves $\omega(p)$ as a function of 
the dimensionless transverse momentum $p$, under the positive average spin polarization 
$S_0 = 1$ and the dimensionless strength of dipolar interactions $\gamma_D = 0.1$, for 
different dimensionless parameters of the quadratic Zeeman effect $\zeta$. The spectrum 
is stable for $\zeta \geq -0.2$ (Fig. 1a) and $\zeta \leq -1.4$ (Fig. 1b).  

\vskip 2cm
{\bf Figure 2}. Dimensionless spectrum of spin waves as a function of the dimensionless
transverse momentum, under the negative spin polarization $S_0=-1$ and $\gm_D = 0.1$, for 
different quadratic Zeeman-effect parameters $\zeta$. The spectrum is stable for either 
$\zeta \geq 1.8$ (Fig. 2a) or $\zeta \leq 0.6$ (Fig. 2b).

\newpage

\begin{figure}[ht]
\centerline{\hbox{
\includegraphics[width=8cm]{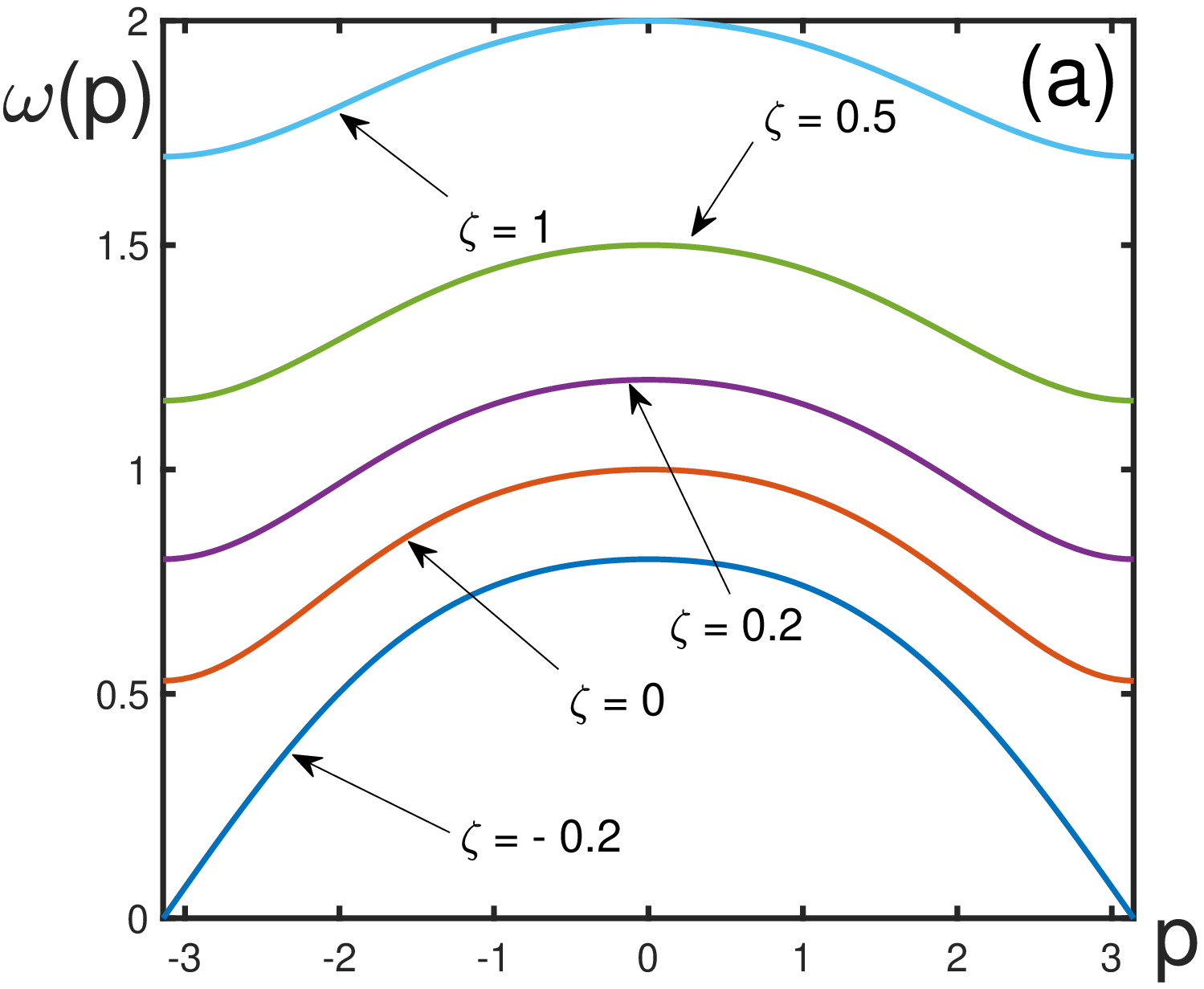} \hspace{1cm}
\includegraphics[width=8cm]{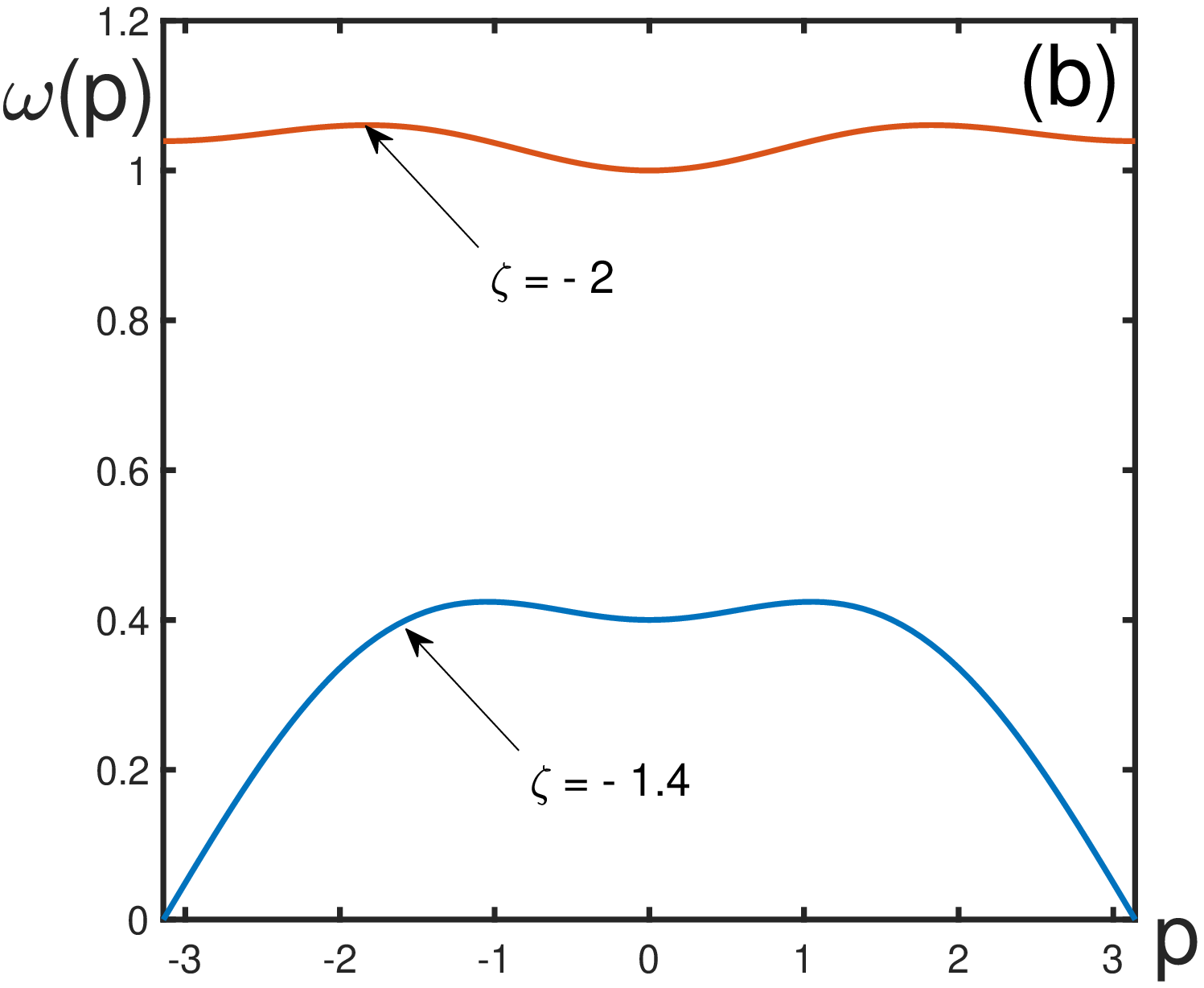} }   }
\caption{Dimensionless spectrum of spin waves $\omega(p)$ as a function of the 
dimensionless transverse momentum $p$, under the positive average spin polarization 
$S_0 = 1$ and the dimensionless strength of dipolar interactions $\gamma_D = 0.1$, 
for different dimensionless parameters of the quadratic Zeeman effect $\zeta$. The 
spectrum is stable for $\zeta \geq -0.2$ (Fig. 1a) and $\zeta \leq -1.4$ (Fig. 1b).  
}
\label{fig:Fig.1}
\end{figure}

\begin{figure}[ht]
\centerline{\hbox{
\includegraphics[width=8cm]{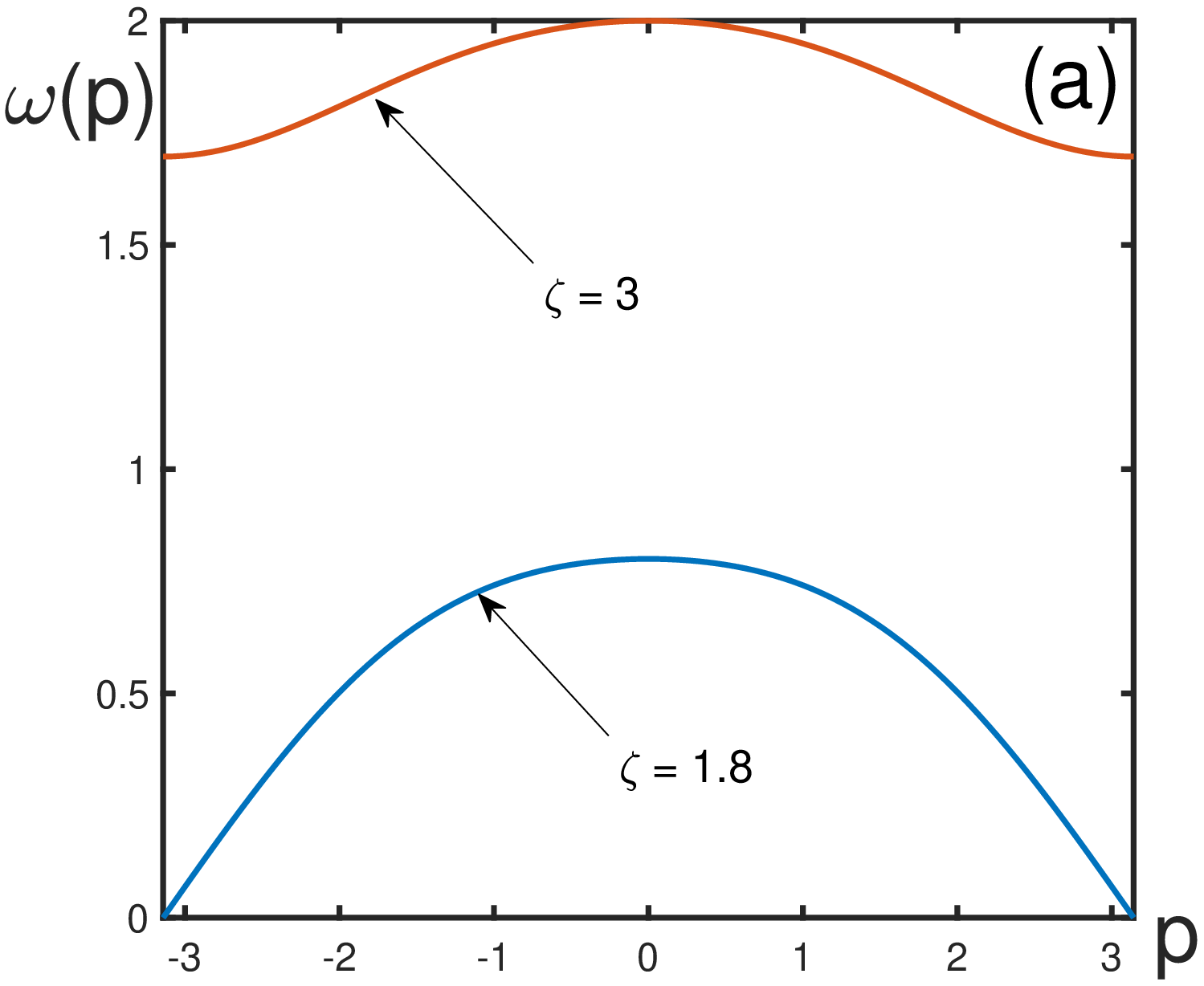} \hspace{1cm}
\includegraphics[width=8cm]{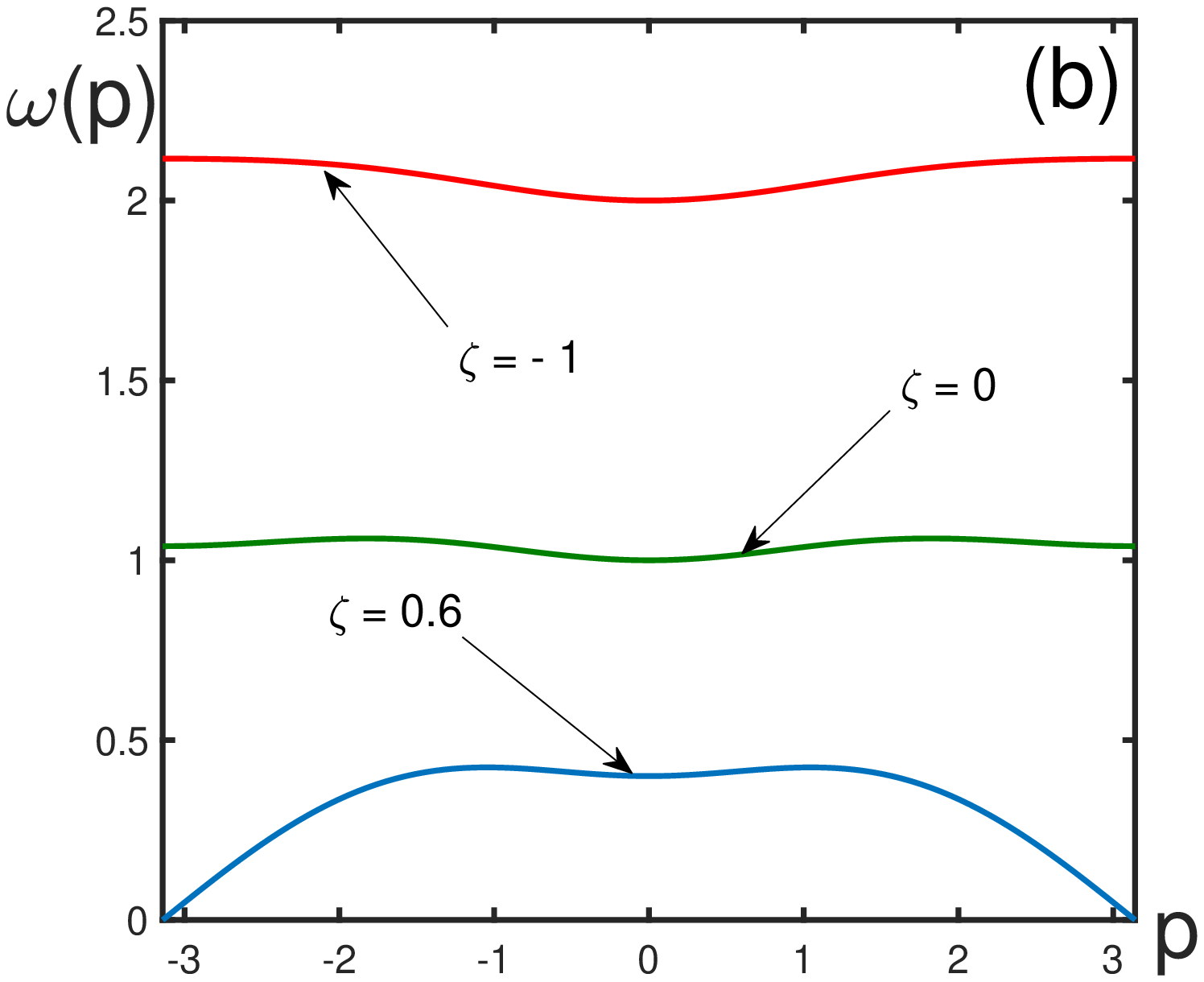} }   }
\caption{Dimensionless spectrum of spin waves as a function of the dimensionless
transverse momentum, under the negative spin polarization $S_0=-1$ and $\gm_D=0.1$, 
for different quadratic Zeeman-effect parameters $\zeta$. The spectrum is stable for 
either $\zeta \geq 1.8$ (Fig. 2a) or $\zeta \leq 0.6$ (Fig. 2b).   
}
\label{fig:Fig.2}
\end{figure}

\end{document}